\documentclass[aps,prl,twocolumn]{revtex4}
\usepackage{graphicx}

\def \be{\begin{equation}}
\def \ee{\end{equation}}
\def \ba{\begin{array}}
\def \ea{\end{array}}
\def \bea{\begin{eqnarray}}
\def \eea{\end{eqnarray}}
\def \nn{\nonumber}

\def \etal{{\it {et al}.}}

\def \a{{\alpha}}

\def \TMTSFPF{(TMTSF)$_2$PF$_6$ }
\def \SOinc {SO(3)$_{\rm spin}\times$SO(4)$_{\rm iso}$ }

\begin{document}

\title{SO(4) Theory of Antiferromagnetism and Superconductivity in Bechgaard Salts}

\author{Daniel Podolsky,  Ehud Altman, Timofey Rostunov, and Eugene Demler}
\address{Department of Physics, Harvard University, Cambridge MA 02138}
\date{\today}

\begin{abstract}
Motivated by recent experiments with Bechgaard salts, we
investigate the competition between antiferromagnetism and triplet
superconductivity in quasi one-dimensional electron systems.  We
unify the two orders in an SO(4) symmetric framework, and
demonstrate the existence of such symmetry in one-dimensional
Luttinger liquids. SO(4) symmetry, which strongly constrains the
phase diagram, can explain coexistence regions between
antiferromagnetic, superconducting, and normal phases, as observed
in (TMTSF)$_2$PF$_6$. We predict a sharp neutron scattering
resonance in superconducting samples.
\end{abstract}

%Motivated by recent experiments with Bechgaard salts, we
%investigate the competition between antiferromagnetism and triplet
%superconductivity in quasi 1D electron systems.  The
%two orders can be unified in an SO(4) symmetric framework. We
%demonstrate the existence of such symmetry in 1D
%Luttinger liquids. SO(4) symmetry requires a first order
%transition between antiferromagnetic and unitary triplet
%superconducting phases, and a weakly first order transition
%between antiferromagnetic and normal phases near the SO(4)
%symmetric point.  This explains coexistence regions between
%superconducting and antiferromagnetic phases, and between
%antiferromagnetic and normal phases observed in (TMTSF)$_2$PF$_6$.
%We predict a sharp resonance mode in inelastic neutron scattering
%experiments in superconducting samples.

\maketitle

%\section{Introduction}

A common feature of many strongly correlated electron systems is
proximity of a superconducting state to some kind of magnetically
ordered insulating state.  Examples include  organic
materials\cite{Jerome1994,Lefebvre2000}, heavy fermion
superconductors\cite{Mathur1998,Kitaoka2001}, and high T$_c$
cuprates\cite{Maple1998}. Several theoretical analyses suggest
that a strong repulsion between the two orders plays an
important role in determining the phase diagram and low energy
properties of these
materials\cite{Scalapino1995}. %,Carlson2002,Altman2002,Sachdev2003}.
The idea of competing orders was developed into the SO(5) theory
of high T$_c$ superconductivity of S.C.~Zhang\cite{Zhang1997}.
SO(5) symmetry has also been applied to study competition of
ferromagnetism and $p$-wave superconductivity in
Sr$_2$RuO$_4$\cite{Murakami1999}, and antiferromagnetism and
$d$-wave superconductivity in
$\kappa$-BEDT-TTF salts\cite{Murakami2000}.

In this paper we consider the interplay of antiferromagnetism (AF)
and triplet superconductivity (TSC) in quasi one-dimensional (Q1D)
electron systems. Our study is motivated by Q1D Bechgaard salts
(TMTSF)$_2$X. The most well studied material from this family,
(TMTSF)$_2$PF$_6$, is an antiferromagnetic insulator at ambient
pressure and a superconductor at high
pressures\cite{Jerome1980,Andres1980,Vuletic2002,Kornilov2003}.
The symmetry of the superconducting order parameter in
(TMTSF)$_2$PF$_6$ is not yet fully established, but there is
strong evidence that electron pairing is spin triplet: the
superconducting T$_c$ is strongly suppressed by
disorder\cite{Choi1984}; %,Tomic1983};
critical magnetic field
$H_{c2}$ in the interchain direction exceeds the paramagnetic
limit\cite{Lee1997}; the electron spin susceptibility, obtained
from Knight shift measurements, does not decrease below
T$_c$\cite{Lee2000}. In another material from this family,
(TMTSF)$_2$ClO$_4$, superconductivity is stable at ambient
pressure and also shows signatures of triplet
pairing\cite{Takigawa1987,Ha2003,Joo2004,Oh2004}.  Insulator to
superconductor transition as a function of pressure has also been
found in(TMTSF)$_2$AsF$_6$\cite{Brusetti1982}.

%\section{SO(4) symmetry in a Luttinger liquid: generators and order parameter}

The phase diagram of interacting electrons in one dimension was
obtained in Ref.~\onlinecite{Giamarchi1989} using bosonization and
renormalization group (RG) analyses. At incommensurate filling,
this system has a phase boundary between spin density wave (SDW)
and TSC phases when $K_\rho=1$ and $g_1>0$ ($K_\rho$ is the
Luttinger parameter in the charge sector, $g_1$ the backward
scattering amplitude). The starting point of our discussion is the
observation that, in the absence of umklapp, 1D Luttinger liquids
have an ``isospin" SO(4)$_{\rm iso}$ symmetry\cite{Efetov1981} at
the boundary between SDW and TSC phases. To define this symmetry,
we introduce the charge of left and right moving electrons,
$Q_\pm$, and two new operators, $\Theta_\pm$, ($r=\pm$)
\begin{eqnarray}
Q_r&=& \frac{1}{2} \sum_{ks} \left( a_{r,ks}^\dagger a_{r,ks}
-\frac{1}{2} \right) \nonumber\\
\Theta_r^\dagger&=&r\sum_k a_{r,k\uparrow}^\dagger
a_{r,-k\downarrow}^\dagger.
\label{incomQuantumGenerators}
\end{eqnarray}
Here $a_{\pm,ks}^\dagger$ creates right/left moving electrons of
momentum $\pm k_f+k$ and spin $s$.  Combining these according to
$J^r_x=(\Theta^\dagger_r+\Theta_r)/2$,
$J^r_y=(\Theta^\dagger_r-\Theta_r)/2i$, and $J^r_z=Q_r$, we see
that the generators satisfy two independent chiral SO(3) algebras,
$\left[ J^r_a, J^{r'}_b
\right]=i\delta^{r,r'}\epsilon^{abc}J^r_a$. The product of left
and right algebras yields the total isospin group SO(4)$_{\rm iso}
\approx$SO(3)$_R\times$SO(3)$_L$.  The Luttinger Hamiltonian at
incommensurate filling, ${\cal H}$, generically commutes with the
charge operators $Q_\pm$. Using a bosonized form of ${\cal H}$, it
can be shown that at $K_\rho=1$, the $\Theta_\pm$ operators also
commute with ${\cal H}$\cite{Efetov1981,Podolsky2004b}. Thus, at
$K_\rho=1$, SO(4)$_{\rm iso}$ forms an exact symmetry of ${\cal
H}$. In addition, for spin-symmetric interactions, which describe
Bechgaard salts to a very good
approximation\cite{Podolsky2004b,Torrance1982}, the system has
SO(3)$_{\rm spin}$ symmetry, $\left[ S_\alpha, S_\beta \right]=i
\epsilon^{\alpha\beta\gamma} S_\gamma,$ generated by the total
spin
\begin{eqnarray}
S_\alpha=\frac{1}{2} \sum_{r,kss'}a_{r,ks}^\dagger \sigma^\a_{ss'}
a_{r,ks'}.
\end{eqnarray}
Hence, for $K_\rho=1$, {\it i.e.} the line separating SDW and TSC phases,
the system has full \SOinc symmetry. We emphasize that for Luttinger liquids
at incommensurate filling, this symmetry always appears at the SDW/TSC phase
boundary and does not require fine-tuning of the parameters.

This symmetry can be used to unify SDW and TSC order parameters.
SDW order away from half filling is described by a complex vector
order parameter,
\begin{eqnarray}
\Phi_\a =\sum_{kss'} a^\dagger_{+,ks} \sigma^\a_{ss'} a_{-,ks'}.
\end{eqnarray}
Q1D band structure restricts the orbital component of triplet
superconducting order to be $\vec{\Psi}(\vec{p})\propto p_x$, with
$x$ the coordinate along the chains.  Thus, the TSC order
parameter is also a complex vector,
\begin{eqnarray}
\Psi_\a^\dagger &=& \frac{1}{i} \sum_{kss'} a^\dagger_{+,ks}
(\sigma^\a \sigma_2)_{ss'} a^\dagger_{-,-ks'} \label{SDWorder}
\end{eqnarray}
The four vector order parameters Re$\vec{\Phi}$, Im$\vec{\Phi}$,
Re$\vec{\Psi}$, and Im$\vec{\Psi}$ can be combined into a
4$\times$3 matrix,
\begin{eqnarray}
P_{\bar{a}\a} &=& \left(\begin{array}{cccc}
({\rm Re} \vec{\Psi})_x & ({\rm Im} \vec{\Psi})_x & ({\rm Re} \vec{\Phi})_x & ({\rm Im} \vec{\Phi})_x\\
({\rm Re} \vec{\Psi})_y & ({\rm Im} \vec{\Psi})_y & ({\rm Re} \vec{\Phi})_y & ({\rm Im} \vec{\Phi})_y\\
({\rm Re} \vec{\Psi})_z & ({\rm Im} \vec{\Psi})_z & ({\rm Re} \vec{\Phi})_z & ({\rm Im} \vec{\Phi})_z\\
\end{array}
\right)
\end{eqnarray}
Each column (row) of $\hat{P}$ transforms independently as a
vector under the action of SO(3)$_{\rm spin}$ (SO(4)$_{\rm iso}$).

The $\Theta_r$ operators are reminiscent of Yang's $\eta$
operator, which generates an SO(4) symmetry for the Hubbard
model\cite{Yang1990}. Unlike $\eta$, whose center of
mass momentum is always the commensurate wave vector $\pi$, the
$\Theta_r$ have their momenta at the wave vectors
$\pm 2 k_f$. This is crucial for defining the symmetry at
arbitrary electron density; in contrast, Yang's SO(4)
applies only at half-filling.

Real materials are only Q1D and coupling between chains gives
rise to finite temperature phase transitions. However, as long as
3D coupling is weaker than the intrachain tunnelling and
interactions, the nature of the ordered state is determined by the
most divergent susceptibility within individual chains. Hence, for
Q1D materials near the SDW/TSC boundary we expect to find
$K_\rho$ close to one, and to find approximate \SOinc symmetry.
Then, the phase diagram in three spatial dimensions is obtained
from a Ginzburg-Landau (GL) free energy whose form is strongly
constrained by symmetry,
\begin{eqnarray}
F &=& \frac{1}{2}  (\nabla P_{\bar{a}\alpha})^2 +
\bar{r}P_{\bar{a}\alpha}^2 +\delta r
\left(P_{1\a}^2+P_{2\a}^2-P_{3\a}^2-P_{4\a}^2\right)\nn\\
&+&\tilde{u}_1 P_{\bar{a}\alpha}^2 P_{\bar{b} \beta}^2
+\tilde{u}_2 P_{\bar{a}\alpha} P_{\bar{a}\beta} P_{\bar{b} \alpha}
P_{\bar{b}\beta}
\label{GinvGL}
\end{eqnarray}
This is the most general expression with SO(3)$_{\rm spin}\times$
SO(4)$_{\rm iso}$ symmetric quartic coefficients. We follow the
common assumption that changing the external control parameters of
the system, e.g. temperature and pressure, only affects the
quadratic coefficients. These are thus allowed to break the
symmetry and tune the phase transition. For $\delta r\ne 0$, the
symmetry is broken down to SO(3)$_{\rm spin}\times$SO(2)$_{\rm
c}\times$SO(2)$_{\rm t}$, where SO(2)$_{\rm c}$ is generated by
the total charge $Q=Q_++Q_-$, and SO(2)$_{\rm t}$ comes from the
lattice translational symmetry\cite{Zhang02}.

At half-filling, umklapp scattering must be taken into
consideration. This is the case for Bechgaard salts, where
structural dimerization splits the conduction band into a full
lower band, and a half-filled upper band.  Umklapp turns two right
movers into left movers, and vice versa,
\begin{eqnarray}
{\cal H}_3 &=& \frac{g_3}{2L}\sum
a_{+,k+qs}^\dagger a_{+,p-qt}^\dagger
a_{-,pt}a_{-,ks}
+ h.c.
\label{umklapp}
\end{eqnarray}
The term (\ref{umklapp}) pins the phase of the SDW, reducing it to
the real N\'eel vector $\vec{N}={\rm Re}\vec{\Phi}$ (for $g_3>0$).
In the presence of umklapp there is still a direct AF/TSC
transition at $K_\rho=1$\cite{Jerome1994}.  On the other hand,
$Q_+$ and $Q_-$ are no longer conserved separately, and
SO(4)$_{\rm iso}$ symmetry of the free energy is broken. However,
to linear order in $g_3$, SO(4)$_{\rm iso}$ is not broken all the
way down to SO(2)$_{\rm c}$. The contribution of ${\cal H}_3$ to
the free energy is $\Delta F=\frac{g_3}{L}\left(({\rm
Re}\vec{\Phi})^2- ({\rm Im}\vec{\Phi})^2\right)$, which preserves
a residual symmetry SO(3)$_{\rm iso}$, given by the diagonal
subgroup of SO(4)$_{\rm
iso}\approx$SO(3)$_R\times$SO(3)$_L$\cite{Podolsky2004b}. The
generators of SO(3)$_{\rm iso}$ are $I_a=J_a^++J_a^-$. Together
with SO(3)$_{\rm spin}$ invariance, the total symmetry defined by
these operators is SO(4)$\approx$SO(3)$_{\rm
spin}\times$SO(3)$_{\rm iso}$.

The justification for considering small $g_3$ is as follows. The
bare value of $g_3$ is proportional to dimerization, which is only
of order of $1\%$ in (TMTSF)$_2$PF$_6$\cite{Thorup}. Furthermore,
the GL free energy depends on the renormalized value $g_{3,\rm
eff}$ at the crossover scale between 1D and 3D physics, which is
even smaller than the bare value of $g_3$.  Umklapp is irrelevant
inside the TSC phase, as well as on the AF/TSC phase boundary.
Even in the AF phase, where umklapp is a relevant perturbation,
$g_3$ flows near zero before diverging, and this divergence may be
cut off by the onset of 3D coupling. Therefore, everywhere close
to the AF/TSC phase boundary, we can assume that $g_{3,\rm eff}$
is small.

SO(4) symmetry unifies AF and TSC orders, which are now combined
into a $3\times 3$ tensor order parameter,
\begin{eqnarray}
Q_{a\a} &=& \left(\begin{array}{ccc}
\,\,({\rm Re} \vec{\Psi})_x\,\,\,\, ({\rm Im} \vec{\Psi})_x\,\,\,\, N_x \,\,\\
\,\,({\rm Re} \vec{\Psi})_y\,\,\,\, ({\rm Im} \vec{\Psi})_y\,\,\,\, N_y \,\,\\
\,\,({\rm Re} \vec{\Psi})_z\,\,\,\, ({\rm Im}
\vec{\Psi})_z\,\,\,\, N_z \,\,
\end{array}\right)
\label{Qtensor}
\end{eqnarray}
The columns (rows) of $\hat{Q}$ transform as a vector under the
spin (isospin) SO(3) algebra,  $ \left[{S_\alpha,Q_{b\beta}}
\right] = i\epsilon^{\alpha\beta\gamma}Q_{b\gamma}$ ($ \left[
{I_a,Q_{b\beta}} \right]= i\epsilon^{abc}Q_{c\beta}$).  In analogy
with (\ref{GinvGL}), the GL free energy near the AF/TSC phase
boundary is
\begin{eqnarray}
F&=&\frac{1}{2}\left(\nabla Q_{a\alpha}\right)^2+
\bar{r} Q_{a\alpha}^2
+\delta r (Q_{z,\alpha}^2
- Q_{x,\alpha}^2 -Q_{y,\alpha}^2)\nn\\
&+& \tilde{u}_1 Q^2_{a\alpha} Q^2_{b\beta}
+\tilde{u}_2  Q_{a\alpha} Q_{a\beta}
Q_{b\alpha} Q_{b\beta}
\label{NearlySO4GLenergy}
\end{eqnarray}
When $\delta r=0$ the model has full SO(4) symmetry.  Away from
this line it only has spin and charge SO(3)$_{\rm
spin}\times$SO(2)$_{\rm c}$ symmetry. A derivation of the GL free
energy for weakly interacting Q1D electrons yields the model
in (\ref{NearlySO4GLenergy}) with $\tilde{u}_1=21 \zeta
(3)/16\pi^2 v_f T^2$ and $\tilde{u}_2 = -7 \zeta (3)/8\pi^2 v_f
T^2 $\cite{Podolsky2004b}.

\begin{figure}
\includegraphics[width=5cm]{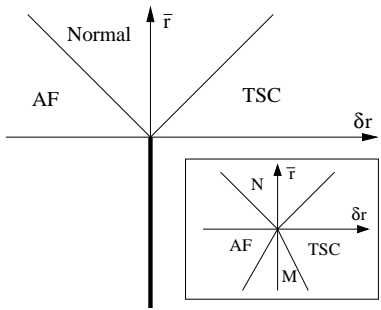}
\caption{Mean field phase diagram of eq.~(\ref{NearlySO4GLenergy})
in the unitary case $\tilde{u}_2<0$. There is a first order
transition (thick line) between AF and TSC phases.  {\bf Inset:}
Corresponding diagram for non-unitary case $\tilde{u}_2>0$. M
denotes a mixed AF/TSC phase. \label{MFdiagram}}
\end{figure}

The properties of model (\ref{NearlySO4GLenergy}) depend strongly
on the sign of $\tilde{u}_2$, which determines whether the triplet
superconductor is unitary ($\Re\vec{\psi}\propto\Im\vec{\psi}$) or
non-unitary ($\Re\vec{\psi}\times\Im\vec{\psi}\ne 0$). We expect
the unitary case, $\tilde{u}_2<0$, to be of experimental relevance
to \TMTSFPF, and in the remainder of this paper we primarily
concentrate on it. The mean field diagram is then composed of an
AF phase separated from a TSC phase by a first order phase
transition, and a disordered (Normal) phase separated from the two
other phases by second order lines (see Fig.~\ref{MFdiagram}).
For completeness, we include in the inset the mean field phase
diagram for the non-unitary case.

\begin{figure}
\includegraphics[width=5cm]{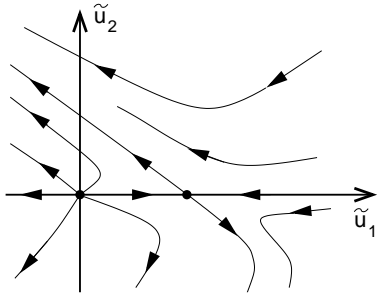}
\caption{Renormalization group flow of the SO(4) symmetric theory
eq.~(\ref{NearlySO4GLenergy}) in $d=4-\epsilon$.  There are no
stable fixed points.  Instead, there are two types of runaway
flow, corresponding to unitary ($\tilde{u}_2<0$) and non-unitary
($\tilde{u}_2>0$) TSC. \label{symflow}}
\end{figure}

To understand the role of thermal fluctuations in model
(\ref{NearlySO4GLenergy}) and in slightly perturbed models where
the quartic coefficients do not lie exactly on the SO(4) symmetric
manifold, we use $4-\epsilon$ RG analysis. We find that the RG
equations have only two fixed points: a trivial Gaussian fixed
point, $\bar{r}=\delta r=\tilde{u}_i=0$, and an SO(9) Heisenberg
point, $\bar{r}\ne 0$, $\delta r=0,$ $\tilde{u}_1\ne 0$,
$\tilde{u_2}=0$. In Fig.~\ref{symflow} we show RG flows in the
SO(4) symmetric plane, where we find runaway flows whenever
$\tilde{u}_2\ne 0$. The analysis can be generalized to order
parameters $\vec{N}$ and $\vec{\Psi}$ that are $N$-component
vectors, in which case the SO(4)$\approx$SO(3)$_{\rm
spin}\times$SO(3)$_{\rm iso}$ symmetry becomes SO($N$)$_{\rm
spin}\times$SO(3)$_{\rm iso}$. We find that even in the large $N$
limit, all flows with $\tilde{u}_2<0$ are runaway flows,
indicating the absence of fixed points with unitary TSC.

\begin{figure}
\includegraphics[width=4.5cm]{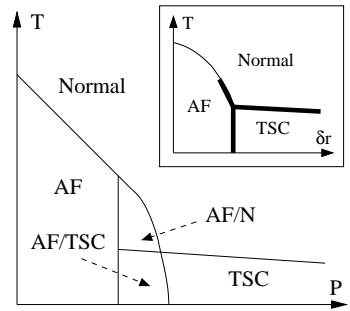}
\caption{Schematic temperature-pressure phase diagram of
\TMTSFPF\cite{Vuletic2002,Kornilov2003}.  AF/N and AF/TSC
correspond to coexistence regimes of the appropriate phases. {\bf
Inset:} Phase diagram for competing AF and unitary TSC states for
model (\ref{NearlySO4GLenergy}) in the large $N$ limit. $\delta r$
tunes the system across the two phases. Thick lines represent
first order transitions. \label{TMTSF_thVSexp}}
\end{figure}

The absence of a fixed point in the RG flow often implies that
fluctuations induce a first order phase transition, thus
precluding a multicritical point in the phase diagram.  In order
to inspect this possibility, we study model
(\ref{NearlySO4GLenergy}) directly in 3D in the large $N$
limit\cite{Podolsky2004b}. The idea of the large $N$ expansion is
to sum self-consistently all bubble diagrams. We find (inset of
Fig.~\ref{TMTSF_thVSexp}) a first order transition between AF and
TSC phases along the SO(4) symmetric line $\delta r=0$, in
agreement with mean field theory. A new feature of the large $N$
limit is the first order transition between the Normal and AF
phases in the vicinity of the critical point.  If we assume that
the experimentally controlled pressure changes an extensive
variable conjugate to $\delta r$, such as the volume of the
system, the first order transition broadens into a coexistence
region of TSC and AF.  This is consistent with the experimental
phase diagram of \TMTSFPF (Fig.~\ref{TMTSF_thVSexp}).
An unusual feature of the theoretical phase diagram (inset,
Fig.~\ref{TMTSF_thVSexp}) is the first order transition between
Normal and TSC phases. This is similar to the fluctuation-driven
first order transition between Normal and superfluid phases
proposed for $^3$He by Bailin \etal\cite{Bailin1977}.  In bulk
$^3$He, the coherence length is very long and the transition is
mean-field like.  The fluctuation region is so small that tiny
discontinuities at the transition caused by fluctuations would be
impossible to observe. In Q1D systems, such as Bechgaard salts,
the fluctuation region is expected to be large\cite{Schulz1981}
and the discontinuous nature of the Normal to TSC transition may
be experimentally accessible. This transition has been
investigated through specific heat measurements in
(TMTSF)$_2$ClO$_4$\cite{Garoche1982}. The results were interpreted
as a mean-field BCS transition, although the amplitude of the
specific heat jump was unusually large.  Since $T_c$ in these
materials is very sensitive to impurities, the extra amplitude in
the specific heat jump might be attributed to broadening due to
disorder of a $\delta$-function peak in specific heat.
Experimental observation of the coexistence region of Normal and
TSC phases, or of hysteresis effects in resistivity measurements
may also be very difficult. Nearly equal strains in the TSC and
the Normal phases and the nearly temperature independent
superconducting transition temperature can make such a coexistence
region very small. On the other hand, it is possible that spin
anisotropy present in real materials, but not included in our
theoretical analysis, stops the runaway RG flows leading to the
first order transition. Another possibility is that one loop
$4-\epsilon$ RG calculations and the large $N$ expansion do not
capture the correct behavior of model (\ref{NearlySO4GLenergy})
for $\epsilon=1$ and $N=3$. For the Normal to unitary TSC
transition, De Prato {\it et al.} argued that a stable fixed point
describing the second order phase transition appears in the
six-loop expansion of the GL free energy\cite{DePrato2002}. We
hope that future experiments will investigate the nature of the
Normal to TSC transition in Bechgaard salts in more detail.

%\section{Collective modes}

The most dramatic consequence of enhanced symmetry at the phase
transition is the prediction of new low energy collective
excitations.  In the vicinity of the AF/TSC phase transition,
SO(4) symmetry leads to a new collective mode, corresponding to
the $\Theta$ operator that rotates AF and TSC orders into each
other\cite{Podolsky2004b}. As pressure is varied toward the phase
transition, breaking of the SO(4) symmetry is reduced, leading to
a decrease in the energy of the $\Theta$ excitation. Mode
softening is not expected generically at a first order phase
transition and identifies the $\Theta$-resonance as a generator of
the SO(4) quantum symmetry. Weak symmetry breaking due to
interchain coupling and higher order umklapp terms may lead to a
small gap and to finite broadening of $\Theta$, even at the AF/TSC
phase boundary. Deep in the normal phase the $\Theta$ excitation
cannot be probed by conventional methods, such as electromagnetic
waves or neutron scattering, as these only couple to particle-hole
channels (e.g. spin or density) and $\Theta$ is a collective mode
in the particle-particle channel. The situation changes when the
system becomes superconducting. In the presence of a condensate of
Cooper pairs, charge is not a good quantum number and
particle-particle and particle-hole channels mix.  The $\Theta$
excitation should thus appear as a resonance in inelastic neutron
scattering experiments\cite{Podolsky2004b}, and its intensity
should be proportional to the square of the superconducting
amplitude $|\vec{\Psi}|^2$.
 For Q1D Bechgaard salts,
%(TMTSF)$_2$PF$_6$,
we expect strong pairing fluctuations even above T$_c$. Hence,
precursors of the $\Theta$ resonance may be visible in the normal
state, with strong enhancement of the resonant scattering
intensity appearing when long range TSC order develops.

It is useful to put the SO(4) model of AF/TSC competition in
Bechgaard salts in the general perspective of electron systems
with competing orders.  In the case of the SO(5) theory of AF and
$d$-wave SC in 2d systems\cite{Zhang1997,Murakami2000}, it is
difficult to construct realistic microscopic models with such
symmetry (see e.g.
Refs.~\onlinecite{Tchernyshyov2001,Henley1998}). By contrast,
SO(4) symmetry in Bechgaard salts arises naturally from a
conventional Luttinger description.
%XXXX
%, and clarify the applicability of the Luttinger model for
%describing these materials.
We also point out that tuning across the AF/TSC phase boundary in
(TMTSF)$_2$PF$_6$ can be done in the same sample by varying
pressure, whereas tuning the AF/SC transition in the cuprates
requires using different samples.  Thus, we consider
(TMTSF)$_2$PF$_6$ a good candidate for experimental observation of
emergence of higher symmetry in a strongly correlated electron
system.

%XXXX
In the discussion above, we assumed Luttinger liquid behavior in
individual chains to motivate the approximate SO(4) symmetry at
the AF/TSC phase boundary.  It has been suggested that for the
superconducting phase of Bechgaard salts, interchain hopping is
strong enough to turn the system into a strongly anisotropic Fermi
liquid\cite{Vescoli1998,Bourbonnais1999}.  The decreased nesting
condition in this case strongly affects antiferromagnetism.  For
the classical symmetry of the GL free energy, this effect can be
absorbed into the normalization of the field $\vec{N}$, so that
the GL parameters only display a small deviation from SO(4)
symmetry at the mean-field level. Thus, we do not expect a
qualitative change in the phase diagram presented in
Fig.~\ref{TMTSF_thVSexp} (see Ref.~\cite{Podolsky2004b} for a
detailed discussion). To verify the approximate quantum SO(4)
symmetry for the strongly anisotropic Fermi liquid, one can study
the spectrum of collective excitations using an RPA-type analysis
and verify the existence of the $\Theta$
excitation\cite{Demler98}. These results will be presented
elsewhere.

In summary, we introduced an SO(4) framework for the competition
between AF and TSC in Q1D electron systems. The microscopic origin
of the SO(4) symmetry at the transition between AF and TSC orders
was identified in the Luttinger liquid model. Our results have
direct implications for Q1D organic superconductors from the
(TMTSF)$_2$X family. For example, first order transitions between
AF and  TSC phases, and between AF and Normal
phases, explain the AF/TSC and the AF/Normal coexistence regions
found in the phase diagram of (TMTSF)$_2$PF$_6$
\cite{Vuletic2002}.  We also argue that the Normal/TSC transition
in these materials could be
weakly first order.  We predict a sharp resonance in neutron
scattering, whose characteristics identify it unambiguously as a
generator of SO(4) symmetry.

We thank S. Brown, P. Chaikin, B.I. Halperin, S. Sachdev,  D.-W.
Wang, and S.C. Zhang for useful discussions. This work was
supported by Harvard NSEC.

%%%%%%%%%%%%%%%%%%%%%%%%%%%%%%%%%%%%%%%%%%%

\end{document}